\documentclass[conference]{IEEEtran}
\IEEEoverridecommandlockouts
\usepackage{cite}
\usepackage[utf8]{inputenc}
\usepackage{amsmath,amssymb,amsfonts}
\usepackage{graphicx}
\usepackage{textcomp}
\usepackage{xcolor}
\def\BibTeX{{\rm B\kern-.05em{\sc i\kern-.025em b}\kern-.08em
    T\kern-.1667em\lower.7ex\hbox{E}\kern-.125emX}}

\usepackage{hyperref}
\usepackage{cleveref}
\usepackage{caption}
\usepackage{subcaption}
\usepackage{paralist}
\usepackage{tabularx,booktabs,longtable}
\usepackage{algorithm}
\usepackage{algorithmic}

\begin{document}

\title{GrapheonRL: A Graph Neural Network and Reinforcement Learning Framework for Constraint and Data-Aware Workflow Mapping and Scheduling in Heterogeneous HPC Systems 

}

%============================
% Commented on 20241211
%============================
\author{\IEEEauthorblockN{1\textsuperscript{st} Aasish Kumar Sharma}
\IEEEauthorblockA{\textit{Faculty of Mathematics and Computer Science} \\
\textit{Georg-August-Universität Göttingen}\\
Göttingen, Germany \\
aasish-kumar.sharma@gwdg.de}
%0000-0002-7514-2340
\and
\IEEEauthorblockN{2\textsuperscript{th} Julian Kunkel}
\IEEEauthorblockA{\textit{Faculty of Mathematics and Computer Science} \\
\textit{Georg-August-Universität Göttingen}\\
Göttingen, Germany \\
julian.kunkel@gwdg.de}
}
%=========================

\maketitle
\thispagestyle{plain}
\pagestyle{plain}

% Main idea, Snakemake applyies workflow analysis and could deploy to SLURM, Kubernets 
% Snakemake applies scheduler ILP (best but slow solution) and Heuristics (approximate but fast solution)
% To improve the situation introducing GNN_RL (best solution and fast near to Heuristics) approach for analyzing workflows.
\section*{Notice}
This work has been accepted for presentation at the IEEE COMPSAC 2025 Conference. © 2025 IEEE. Personal use of this material is permitted. The final published version will be available via IEEE Xplore at: \url{https://ieeexplore.ieee.org/}
\vspace{1em}

\begin{abstract}
Effective resource utilization and decreased makespan in heterogeneous High Performance Computing (HPC) environments are key benefits of workload mapping and scheduling. Tools such as Snakemake, a workflow management solution, employs Integer Linear Programming (ILP) and heuristic techniques to deploy workflows in various HPC environments like SLURM (Simple Linux Utility for Resource Management) or Kubernetes. Its scheduler factors in workflow task dependencies, resource requirements, and individual task data sizes before system deployment. ILP offers optimal solutions respecting constraints, but only for smaller workflows. Meanwhile, meta-heuristics and Heuristics offer faster, though suboptimal, makespan. As problem sizes, system constraints, and complexities evolve, maintaining these schedulers becomes challenging. 

In this study, we propose a novel solution that integrates Graph Neural Network (GNN) and Reinforcement Learning (RL) to flexibly handle workflows, dynamic constraints and heterogeneous resources while providing quick responses. GNN manages dependencies and resource requirements, and RL optimizes scheduling decision-making via a learned policy, overcoming the need for a comprehensive global search. Experimental results with different datasets demonstrate that this method effectively adapts to different workflows, adheres to HPC constraints, and offers optimal solutions akin to ILP but with drastically reduced execution times ($76\%$ faster), comparable to heuristic methods (only $3.85\times$ slower than OLB). Our contribution is to provide a robust yet scalable mapping and scheduling solution that can handle changing constraints, as well as workload sizes and complexities in a Heterogeneous HPC Compute Continuum system landscape.

\end{abstract}

\begin{IEEEkeywords}
Heterogeneous High Performance Computing, Mapping and Scheduling, Graph Neural Networks, Reinforcement Learning, Mixed Integer Linear Programming, Heuristics, Snakemake, SLURM, Kubernetes, Artificial Intelligence, Machine Learning 
\end{IEEEkeywords}

\section{Introduction}
\label{sec:Introduction}

Heterogeneous High-Performance Computing (HPC) systems integrate diverse resources like CPUs, GPUs, and specialized accelerators to manage workloads from IoT (Internet of Things), edge, cloud, and HPC domains. Effective automation is crucial for optimizing resource use, minimizing makespan, and increasing throughput. Existing tools, such as SLURM \cite{yoo2003slurm}, Kubernetes \cite{burns2019kubernetes}, and Snakemake \cite{koster2012snakemake}, encounter limitations due to static, greedy resource allocation and challenges in dynamic environments, highlighting the need for advanced scheduling and management protocols.

\subsection{Motivation and Challenges}

Key challenges in heterogeneous HPC systems include:

\textit{Complex Dependencies} - Managing intricate task inter-dependencies requires advanced scheduling.
\textit{Resource Heterogeneity} - Adaptive policies are needed for effectively utilizing diverse architectures.
\textit{Dynamic Constraints} - Variable constraints necessitate solutions beyond static approaches.

We propose \emph{GrapheonRL}, an AI/ML approach using Graph Neural Networks (GNN) and Reinforcement Learning (RL) to improve workload mapping and scheduling, preserving constraints and reducing search space after training.

This paper is organized as follows: \Cref{sec:Introduction} introduces the research problem and motivation; \Cref{sec:Background} outlines system and workload characteristics; \Cref{sec:RelatedWork} reviews relevant literature; \Cref{sec:Methodology} describes the techniques and system modeling employed; \Cref{sec:ExperimentalSetup} details experimental setups; \Cref{sec:ResultsAndDiscussion} discusses results; finally, \Cref{sec:ConclusionAndFutureWork} provides conclusions and future work recommendations.

\section{Background}
\label{sec:Background}

\subsection{HPC-Compute Continuum (HPC-CC) System}

\begin{figure*}%[H]
    \centering
    \includegraphics[width=\textwidth]{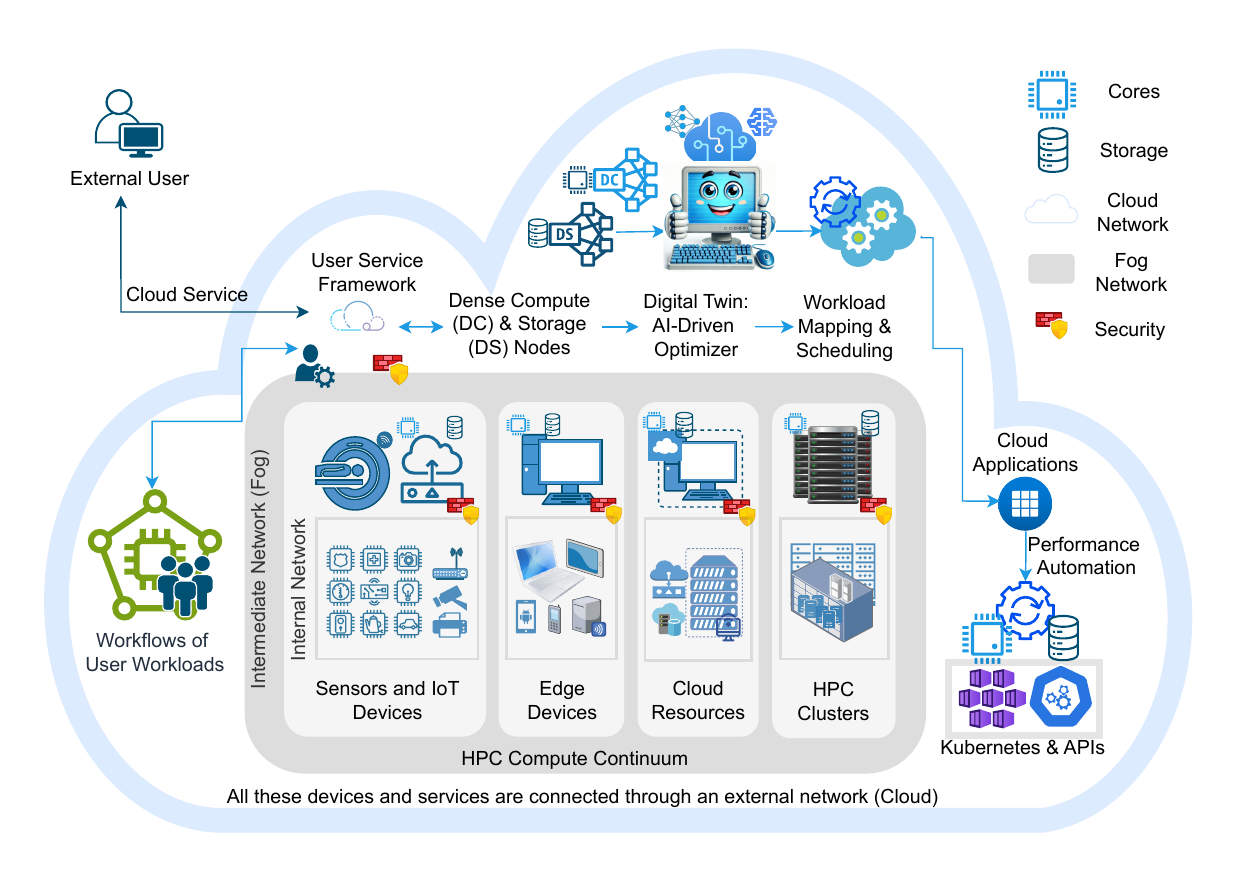}
    \caption{Platform as a Service (PaaS): Digital Twin, User Cloud Service.}
    \label{fig:ComputeContinuumWorkflowOptimizationFramework}
    %\vspace{-1.5em}
\end{figure*}

Modern HPC systems operate within a "compute continuum" (HPC-CC as shown in \Cref{fig:ComputeContinuumWorkflowOptimizationFramework}) encompassing edge, cloud, and supercomputers \cite{dongarra2022hpc}. They employ massive parallelism to solve computationally intensive problems, posing optimization challenges in resource management and task scheduling.

\subsection{Workloads: Workflows}
\label{ssec:WorkloadsWorkflows}

Workloads in the HPC-CC include complex tasks and simulations. For instance, MRI data processing involves capturing data with IoT devices, noise reduction at the edge, secure storage and analysis in the cloud, and executing advanced tasks on HPC systems. \Cref{fig:PetriNetWithResourceAndFeatureRequest}illustrates the workflows for MRI use case and \Cref{fig:ComputeContinuumWorkflowOptimizationFramework}illustrates the overall scenario for analyzing of workflow request and mapping and scheduling them to the system nodes across the continuum.

\subsection{Existing Tools and Limitations}
To manage such workloads for such heterogeneous system we found following key tools: (a) \textit{SLURM:} Manages HPC clusters with static, greedy resource allocation. (b) \textit{Kubernetes:} Orchestrates cloud containers but lacks fine control for HPC workloads. (c) \textit{Snakemake:} Provides workflow management but does require mechanism for resource mapping and scheduling based on remote host resources. Besides, snakemake prioritize tasks within a workflow, deploying them across HPC environments using plugins for SLURM and Kubernetes. Snakemake utilizes methodologies like Integer Linear Programming (ILP) and greedy heuristic techniques to efficiently allocate resources based on task requirements. ILP provides an accurate model for mapping and scheduling problems, adhering to constraints and optimizing objective functions. However, it can become computationally infeasible in large-scale or dynamically changing environments. In these scenarios, Snakemake defaults to greedy heuristics, which may not fully satisfy all constraints and objectives, particularly when ignoring remote system resources.
To address these gaps it needs to: (a) Develop models incorporating actual resources for optimizing resource allocation, task dependencies, and data transfer costs. (b) Create adaptable mapping and scheduling techniques to automate the process. (c) Ensure optimal solutions for large-scale data with reduced complexity.
This study aims to automate and enhance system performance in heterogeneous HPC environments through efficient workload management.

\subsection{Graph Neural Networks (GNN)}
Graph Neural Networks (GNNs) are a class of neural networks specifically designed to process data structured as graphs. They leverage the relational inductive bias of graphs to model the dependencies of data points represented as nodes and edges in a graph (Scarselli et al., 2009). GNNs have been used in various domains, including social networks, molecular chemistry, and computer vision, to perform tasks such as node classification, link prediction, and graph classification \cite{zhou2020graph}.

\subsection{Reinforcement Learning (RL)}
Reinforcement Learning (RL) is an area of machine learning concerned with how agents ought to take actions in an environment to maximize cumulative reward. It contrasts with supervised learning in that the correct input/output pairs are never presented, nor are sub-optimal actions explicitly corrected. Instead, the focus is on finding a balance between exploration (of uncharted territory) and exploitation (of current knowledge) \cite{sutton2018reinforcement}. RL has been successfully applied in areas such as robotics, game playing, and autonomous vehicle navigation.

\section{Related Work}
\label{sec:RelatedWork}

\subsection{Classification of Scheduling Techniques}
Scheduling approaches fall into four primary categories: mathematical modeling, heuristics, meta-heuristics, and machine learning (ML) \cite{AhmadContainerSchedulingTechniques2022}.

\subsubsection{Linear Programming (ILP, MILP)}
Mathematical models, such as ILP and Mixed-Integer Linear Programming (MILP), provide exact solutions to scheduling problems by formulating them as constrained optimization tasks \cite{ILPContainerOrientedJobSchedulingZhang2017, bertsimas1997introduction}. MILP excels in solving small to medium-scale problems optimally but struggles with scalability in large HPC workflows due to exponential solver time \cite{bixby2012solvers, pinedo2016scheduling}.

\subsubsection{Heuristic and Meta-Heuristic Methods}
Heuristics such as Opportunistic Load Balancing (OLB), Minimum Execution Time (MET), and Heterogeneous Earliest Finish Time (HEFT) offer fast, rule-based scheduling \cite{mishra2020LoadBalancingInCloudComputing, liao2014heuristic}. More advanced meta-heuristic techniques, including genetic algorithms and ant colony optimization, provide near-optimal solutions but require careful tuning \cite{blum2003metaheuristics, xuMMultiplePriorityQueueingGeneticAlgorithm2012}. While heuristics are computationally efficient, they often fail to adapt to complex dependencies and heterogeneous workloads.

\subsubsection{AI and Machine Learning Approaches}
Recent advancements in AI-driven scheduling leverage RL to optimize workload placement dynamically \cite{gudeep2025, mangalampalliefficient2024}. Hybrid approaches combining ML with traditional methods demonstrate improved efficiency and adaptability in HPC scheduling \cite{AhmadContainerSchedulingTechniques2022}.

\subsubsection{Graph Neural Networks (GNNs)}
GNNs extend deep learning to structured data, making them suitable for modeling task dependencies and resource allocation in scheduling problems \cite{scarselliGraphNeuralNetwork2009}. They leverage message passing to encode task relationships, facilitating informed decision-making \cite{gilmer2017neural, kipf2017semisupervised, velivckovic2018graph}.

\subsubsection{GNN-RL for HPC Scheduling}
Integrating GNNs with RL enables structured decision-making for dynamic scheduling \cite{zeng2020graph, pan2021gnn}. RL frameworks such as Deep Q-Networks and Policy Gradient methods optimize scheduling policies while leveraging GNN embeddings to capture workload dependencies \cite{mnih2013playing, suttonReinforcementLearningIntroduction2018}. This approach significantly enhances scheduling efficiency in large-scale HPC environments.

\subsubsection{Emerging Trends in GNN-RL for HPC}
GNN-RL models are gaining traction for optimizing HPC workloads, enabling intelligent task placement and resource allocation in dynamic environments \cite{adimora2024gnnrlnodate, zhoudeep2024, songreinforcement2023, yu2022gnnrl, yu2021spatl, yu2021agmc}.

\section{Methodology}
\label{sec:Methodology}

This section outlines the theoretical formulation of our system and workload models, establishing a basis for practical HPC mapping and scheduling applications.

For this we apply experimental modeling approach that serves as a base for systematic data acquisition through controlled trials and allows for model validation via empirical observation and quantification. It is instrumental in corroborating theoretical paradigms, empirically testing hypotheses, and fine-tuning model parameters using real-world datasets. Additionally, optimization protocols are integrated throughout the model development phase, enhancing our ability to forecast workload completion times and optimize resource allocation within an HPC-CC context.

Our automated procedure for workload assignment and scheduling in the HPC-CC is depicted in Figure~\ref{fig:ComputeContinuumWorkflowOptimizationFramework}. It initiates with a user-service framework permitting external user workload submissions. At predetermined intervals, the HPC-CC framework evaluates for new or pending workflows, adhering to the following protocol:

The preliminary phase involves acquiring input parameters of the HPC-CC system. The digital twin, a supervisory component, delivers comprehensive system and workload performance metrics. In scenarios lacking initial performance data (such as the first execution), a theoretical default seed value is utilized. All collected system and workload characteristics are encoded in JSON format, as outlined in \Cref{sssec:SystemAndWorkloadInputFormat}.

Subsequently, the JSON inputs are processed by the workload analyzer. This solver integrates the data with mathematical models detailed in \Cref{ssec:SystemModeling} and evaluates criteria such as resource availability and workflow stipulations. The solver endeavors to optimally map workflows to system resources, while accounting for attributes, limitations, and intended goals.

Upon processing, the solver outputs an ordered list of workflows and tasks, encapsulated in a JSON file containing comprehensive mapping and scheduling data, which is then relayed to the executor.

The executor deploys the workflows within suitable SLURM and Kubernetes cluster queues. Post execution, metrics and logs are retrieved by the monitoring component, facilitating the update of node properties for future executions.

This project framework operationalizes these procedures for experimental evaluation. Our primary emphasis remains on developing models for system and workload mapping and scheduling, hence the monitoring component is not extensively discussed.

%\subsection{System and Workload Modeling Details} 
%\label{ssec:SystemAndWorkloadModelinDetai} 
\subsection{System Modeling}
\label{ssec:SystemModeling}

Following \cite{eadlineHighPerformanceComputing2009}, an HPC system is modeled as a set of clusters, $C$, where each cluster contains multiple nodes, $N$, interconnected via high-speed networks. Each node is defined by a tuple:
\[
N = \{R, F, P\},
\]
where, \textbf{$R: R^r_N$ (Resources):} Quantifiable components such as CPU cores, memory, and I/O bandwidth, where '$r$' is the instance of resources  ($r\in R$), of node $N$. \textbf{$F:F^f_N$ (Features):} Infrastructure-specific capabilities like ISA type or GPU support, where '$f$' is the instance of features ($f\in F$), of node $N$. \textbf{$P:P^p_N$ (Properties):} Performance metrics including processing speed and storage throughput, where '$p$' is the instance of properties ($p\in P$),  of $N$.\newline

\begin{figure}
    \centering
    \includegraphics[width=0.92\linewidth]{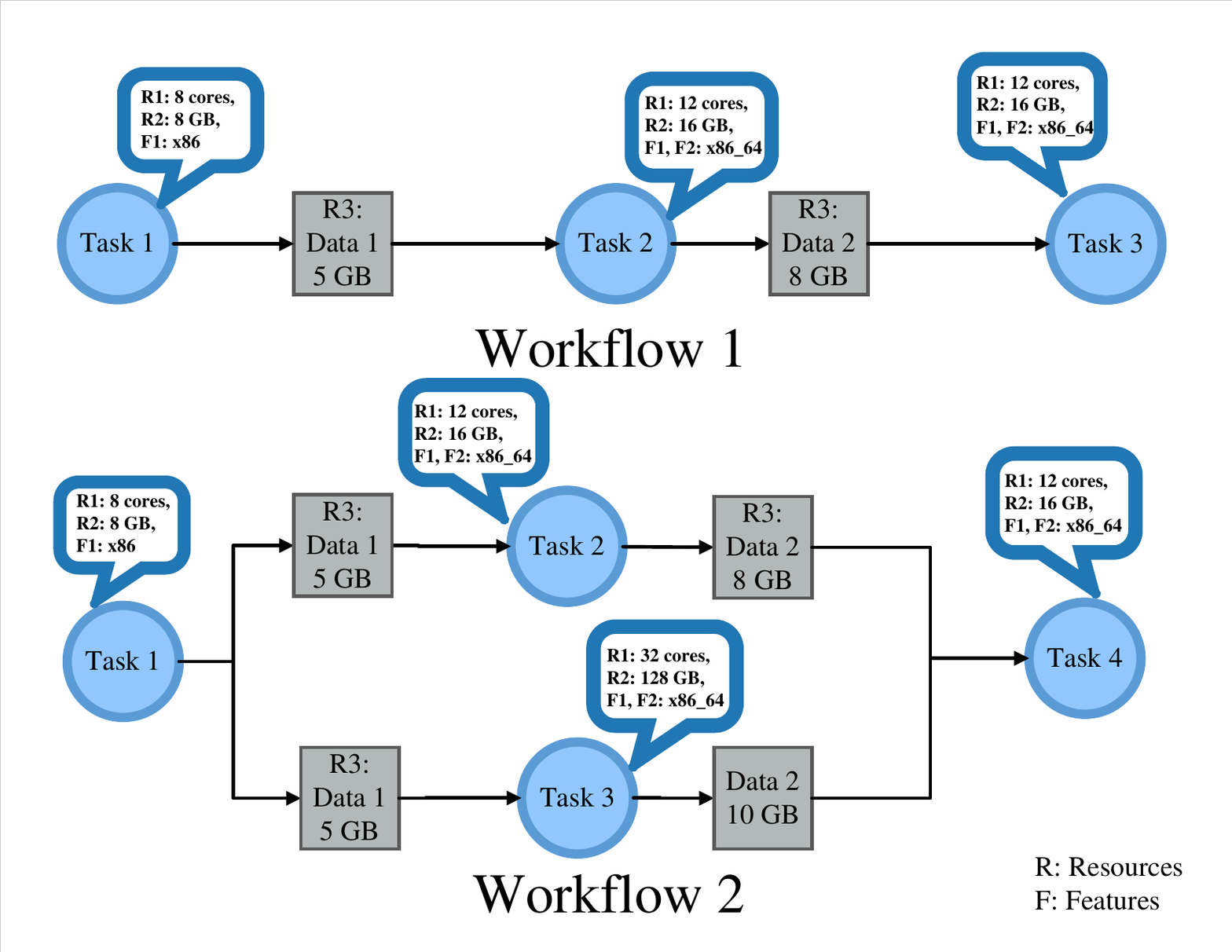}
    \caption{Showing Petri net  example workflows with compute resource and feature requests along with respective intermediate data.}
    \label{fig:PetriNetWithResourceAndFeatureRequest}
    
\end{figure}

Detailed definitions for these variables are provided in Tables~\ref{tab:CombinedNodeInfo}, and \ref{tab:SystemAndWorkloadDataFormat}. Figure~\ref{fig:PetriNetWithResourceAndFeatureRequest} illustrates an example workflow using a Petri net representation that highlights resource and feature requests alongside intermediate data.

\begin{table}%[ht]
    \small
    \centering
    \caption{Node Resources, Features, and Properties}
    \label{tab:CombinedNodeInfo}
    \begin{tabular}{p{1em} p{4em} p{19em}}
    \hline
    \textbf{S.N.} & \textbf{Expression} & \textbf{Definition} \\
    \hline
    \multicolumn{3}{l}{\textbf{Node Resources}} \\
    1. & $R^1_N$ & Number of cores. \\
    2. & $R^2_N$ & Memory capacity (GB). \\
    3. & $R^3_N$ & Storage capacity (GB). \\
    \hline
    \multicolumn{3}{l}{\textbf{Node Features}} \\
    4. & $F^1_N$ & Processor ISA: x86 (CPU). \\
    5. & $F^2_N$ & Processor ISA: x64 (GPU). \\
    6. & $F^3_N$ & Memory type: DDR4. \\
    7. & $F^4_N$ & Memory type: DDR5. \\
    8. & $F^5_N$ & Storage type: HDD. \\
    9. & $F^6_N$ & Storage type: SSD. \\
    10. & $F^7_N$ & Network: Omni-Path. \\
    11. & $F^8_N$ & Network: InfiniBand. \\
    \hline
    \multicolumn{3}{l}{\textbf{Node Properties}} \\
    12. & $P^1_N$ & Processing Speed (e.g., Processor base clock). \\
    13. & $P^2_N$ & Processing Speed (e.g., FLOP/s). \\
    14. & $P^3_N$ & Data Transfer Rate (e.g., PCI/Bus). \\
    \hline
    \end{tabular}
    \vspace{-1em}
\end{table}

\begin{table}%[ht]
    \centering
    \small
    \caption{System and Workload Data Format}
    \label{tab:SystemAndWorkloadDataFormat}
    \begin{tabular}{rp{10em}p{12em}}
        \hline
        \textbf{S.N.} & \textbf{Headers} & \textbf{Details}  \\% & \textbf{Sample Data}
        \hline

        1. & $node$ & $N_{i}$  \\
        2. & $resources$ & $R^r_{i}:\{R^1\}$ cores   \\
        3. & $features$ & $F^f_{i}:\{F^1,...,F^8\}$  \\
        4. & $processing\_speed$ & $P^p_{i}:\{P^1\}$ \\
        5. & $data_transfer\_rate$ & $P^p_{i}:\{P^3\}$  \\

        \hline

        1. & $task$ & $T_{j}$  \\
        2. & $resources$ & $R^r_{j}:\{R^1\}$ cores \\
        3. & $features$ & $F^f_{j}:\{F^1,...,F^8\}$ \\
        4. & $data$ & $R^r_{j}:\{R^3\}$ \\
        5. & $duration$ & $d_{ij}$ \\
        6. & $dependency$ & $\delta_{j,j'}$ \\
        \hline
    \end{tabular}
    \vspace{-1em}
\end{table}

\subsection{Workload Modeling}
\label{ssec:WorkloadModeling}

Workloads are modeled as collections of workflows, each comprising multiple tasks with defined dependencies. In systems such as SLURM, a workload consists of a set of jobs where each job contains one or more tasks; these tasks are interrelated to form a directed acyclic graph (DAG) \cite{vanderaalstWorkflowManagementModels2002}. Formally, a workflow DAG is represented as:
\[
\text{DAG}(\text{Workflow}) = \{V, E\},
\]
where $V$ is the set of tasks and $E$ represents dependencies (e.g., an edge $T_{j'} \rightarrow T_{j}$ indicates that task $T_{j}$ depends on task $T_{j'}$). Each task is defined as a tuple:
\[
T = \{R, F, U, \delta\},
\]
where, \textbf{$R$:} Requested resources ($R^r_T$),
\textbf{$F$:} Required features ($F^f_T$),
\textbf{$U$:} Resource usage ($U^r_T$),
\textbf{$\delta$:} Dependencies among tasks.\newline

The system and workload model variables are summarized in Table~\ref{tab:SystemAndWorkloadDataFormat}. Analysis of task dependencies, including data transfers, is used to estimate inter-node transfer times, while resource usage is evaluated based on performance characteristics (properties) from Table~\ref{tab:CombinedNodeInfo}.

\subsubsection{Mapping and Scheduling Techniques}
During study we found, the Snakemake scheduler initially employs an ILP method and, if no feasible solution is found within 10 seconds, reverts to a greedy heuristic \cite{koster2012snakemake} as illustrated in \Cref{alg:SnakemakeAlgorithm}. 
\begin{algorithm}
  \caption{Snakemake ILP-Greedy Scheduler}
  \begin{algorithmic}[1] 
  %\small
    \REQUIRE Workflow DAG, Resource Limits, Job Requirements
    \ENSURE Job Schedule, Updated Resource Pool
    
    \STATE Initialize global resources.
    \STATE Create a queue of ready jobs. 
    \WHILE{there are ready jobs}
      \STATE Select a ready job from the queue. 
      \IF{sufficient resources are available}
        \STATE Allocate resources to the job.
        \STATE Submit the job for execution.
      \ELSE
        \STATE Return the job to the waiting queue.
      \ENDIF
    \ENDWHILE
    \STATE \RETURN Job Schedule, Updated Resource Pool
  \end{algorithmic}
  \label{alg:SnakemakeAlgorithm}
\end{algorithm}
However, Snakemake estimates resource usage based on local host resources, its default settings may not adequately reflect remote environments (e.g., SLURM or Kubernetes) or host-specific priorities, necessitating modifications to incorporate updated constraints and objectives. Therefore, we extend and design our own scheduler applying MILP and Heuristics approaches as per our objectives and constraints.

\subsection{MILP Algorithm With Required Constraints and Objectives}
We employ a MILP approach, implemented using Python's PuLP library, to solve the resource allocation and scheduling problem as per our constraints. \Cref{alg:MILPAlgorithm} details the algorithm, which incorporates task dependencies, DTT, and resource constraints. The objective is to minimize the weighted sum of resource usage and makespan, subject to assignment, resource, feature, timing, and dependency constraints.

\begin{algorithm}
\caption{MILP: Resource Allocation and Scheduling with DTT}
\begin{algorithmic}[1]
%\small
\STATE \textbf{Input:} 
\STATE \hspace{1em} Nodes: $\{(R_i, F_i)\}_{i \in |N|}$ \hfill \# Resources and features of each node
\STATE \hspace{1em} Tasks: $\{(R_j, F_j, d_j, \delta_j)\}_{j \in |T|}$ \hfill \# Resources, features, duration, and dependencies of each task
\STATE \hspace{1em} Data Transfer Time: $\{d_{jj'}\}_{(j,j') \in \delta}$ \hfill \# Data transfer times between tasks

\STATE \textbf{Variables:} 
\STATE \hspace{1em} $x_{ij} \in \{0, 1\}$ \hfill \# Binary variable indicating if task $j$ is assigned to node $i$
\STATE \hspace{1em} $s_j \in \mathbb{R}^+$ \hfill \# Start time of task $j$
\STATE \hspace{1em} $f_j \in \mathbb{R}^+$ \hfill \# Finish time of task $j$
\STATE \hspace{1em} $C_{\text{max}} \in \mathbb{R}^+$ \hfill \# Maximum finish time of all tasks
\STATE \hspace{1em} $y_{jj'} \in \{0, 1\}$ \hfill \# Binary variable for data transfer constraint

\STATE \textbf{Objective:}
\STATE \hspace{1em} Minimize: $ \alpha \cdot \sum_{j \in |T|} \sum_{i \in |N|} U_j \cdot x_{ij} + \beta \cdot  C_{\text{max}}$

\STATE \textbf{Constraints:}
\STATE \hspace{1em} \textbf{Assignment Constraints:}
\FOR{each task $j \in |T|$}
    \STATE $\sum_{i \in |N|} x_{ij} = 1$ \hfill \# Each task must be assigned to exactly one node
\ENDFOR

\STATE \hspace{1em} \textbf{Resource Constraints:}
\FOR{each node $i \in |N|$}
    \STATE $\sum_{j \in |T|} U_j \cdot x_{ij} \leq R_i$ \hfill \# Node resource capacity
\ENDFOR

\STATE \hspace{1em} \textbf{Feature Constraints:}
\FOR{each node $i \in |N|$}
    \STATE $\sum_{j \in |T|} x_{ij} = 1_{F^f_j \subseteq F^f_i}$ \hfill \# Node with requested features
\ENDFOR

\STATE \hspace{1em} \textbf{Timing Constraints:}
\FOR{each task $j \in |T|$}
    \STATE $f_j = s_j + d_j$ \hfill \# Finish time is the start time plus the duration of the task
\ENDFOR

\STATE \hspace{1em} \textbf{Makespan Constraint:}
\FOR{each task $j \in |T|$}
    \STATE $C_{\text{max}} \geq f_j$ \hfill \# Makespan is the maximum finish time of all tasks
\ENDFOR

\STATE \hspace{1em} \textbf{Dependency Constraints:}
\FOR{each pair of tasks $(j, j') \in \delta$}
    \STATE $s_{j'} \geq f_j + d_{jj'} \cdot (1 - y_{jj'})$ \hfill \# Task $j'$ starts after $j$ finishes, considering data transfer time
    \STATE \textbf{Ensure Data Transfer Variable:}
    \STATE $y_{jj'} \geq x_{ij} + x_{i'j'} - 1$ \hfill \# Ensure $y_{jj'}$ is 1 if tasks are assigned to different nodes
\ENDFOR

\STATE \textbf{Solve the MILP:}
\STATE Use a MILP solver to solve the problem

\STATE \textbf{Output:} Optimal task assignments, start times, finish times, and makespan

\end{algorithmic}
\label{alg:MILPAlgorithm}
\end{algorithm}

\subsection{Heuristic Algorithm}
\subsubsection{HEFT (Heterogeneous Earliest Finish Time)}
HEFT is a well-known heuristic for scheduling HPC tasks. In our simplified implementation (\Cref{alg:HEFTAlgorithm}), HEFT computes an approximate rank for each task (communication cost calculations), sorts the tasks in descending order of rank, and assigns each task to the node that minimizes its finish time—ensuring that each task is scheduled only after its predecessors have completed. This version assumes only basic feasibility checks (e.g., simple resource and feature matching) with validation or explicit data transfer modeling.

\begin{algorithm}
\caption{HEFT (Heterogeneous Earliest Finish Time)}
\label{alg:HEFTAlgorithm}
\begin{algorithmic}[1]
\STATE \textbf{Inputs:} Set of tasks $T=\{t_1,\dots,t_m\}$ with dependencies, set of nodes $N=\{n_1,\dots,n_k\}$
\STATE \textbf{Outputs:} Task-to-node assignment, estimated makespan
\STATE Initialize: $assignment \gets \emptyset$, $node\_schedule[n_j] \gets 0.0$ for all $n_j \in N$

\STATE Compute rank for each task $t \in T$:
\FOR{each task $t \in T$}
    \IF{rank$(t)$ already computed}
        \STATE continue
    \ENDIF
    \STATE $avg\_comp(t) \gets \frac{1}{k} \sum_{n_j \in N} \left( \frac{comp\_time(t)}{speed(n_j)} \right)$
    \IF{$t$ has no dependencies}
        \STATE rank$(t) \gets avg\_comp(t)$
    \ELSE
        \STATE rank$(t) \gets avg\_comp(t) + \max$ of ranks of its dependencies
    \ENDIF
\ENDFOR

\STATE Sort tasks in descending order of rank$(t)$ into list $sorted$

\FOR{each task $t$ in $sorted$}
    \STATE $earliest\_finish \gets \infty$
    \STATE $best\_node \gets$ None
    \FOR{each node $n_j \in N$}
        \STATE $dep\_end \gets \max$ of end times of dependencies of $t$
        \STATE $start\_time \gets \max(node\_schedule[n_j], dep\_end)$
        \STATE $finish\_time \gets start\_time + \frac{comp\_time(t)}{speed(n_j)}$
        \IF{$finish\_time < earliest\_finish$}
            \STATE $earliest\_finish \gets finish\_time$
            \STATE $best\_node \gets n_j$
        \ENDIF
    \ENDFOR
    \STATE $assignment[t] \gets best\_node$
    \STATE $node\_schedule[best\_node] \gets earliest\_finish$
\ENDFOR

\STATE $makespan \gets \max$ finish time among all tasks
\STATE \textbf{return} ($assignment$, $makespan$)
\end{algorithmic}
\end{algorithm}

\subsubsection{OLB (Opportunistic Load Balancing)}
OLB (\Cref{alg:OLBAlgorithm}) schedules tasks based on the current load. It first performs a topological sort to respect task dependencies, then allocates each task to the node with the lowest load (i.e., the earliest available finish time). Node finish times are updated as tasks are assigned, ensuring that subsequent tasks begin only after the node becomes free and all dependency constraints are met.

\begin{algorithm}%[H]
\caption{OLB (Opportunistic Load Balancing)}
\begin{algorithmic}[1]
%\small
\STATE \textbf{Inputs:} Set of tasks $T=\{t_1,\dots,t_m\}$, set of nodes $N=\{n_1,\dots,n_k\}$.
\STATE \textbf{Outputs:} Assignment of tasks to nodes, estimated makespan.

\STATE $node\_finish\_time \leftarrow \{n_j : 0.0 \mid \forall n_j \in N\}$ 
\COMMENT{Tracks current finishing time (or load) on each node}
\STATE $assignment \leftarrow \varnothing$

\STATE \textbf{Compute topological order} of tasks for dependencies: $ordered \leftarrow TopologicalSort(T)$
\FOR{\textbf{each} $t_i$ \textbf{in} $ordered$}
    \STATE $best\_node \leftarrow \text{None}$
    \STATE $best\_finish \leftarrow \infty$
    \FOR{\textbf{each} $n_j \in N$}
        \STATE $dep\_end \leftarrow \max(\text{end\_time of all dependencies of }t_i)$
        \COMMENT{Earliest possible start after deps finish}
        \STATE $start\_time \leftarrow \max(node\_finish\_time[n_j], dep\_end)$
        \STATE $finish\_time \leftarrow start\_time + duration(t_i)$
        \IF{$finish\_time < best\_finish$}
            \STATE $best\_finish \leftarrow finish\_time$
            \STATE $best\_node \leftarrow n_j$
        \ENDIF
    \ENDFOR
    \STATE $assignment[t_i] \leftarrow best\_node$
    \STATE Update $node\_finish\_time[best\_node]$ to $best\_finish$
\ENDFOR

\STATE $makespan \leftarrow \max\limits_{t_i}( finish\_time(t_i) )$
\STATE \textbf{return} ($assignment, makespan$)

\end{algorithmic}
\label{alg:OLBAlgorithm}
\end{algorithm}

\subsection{GNN-RL for Workload Mapping and Scheduling}

Our architecture combines GNNs and RL to optimize HPC task scheduling of a heterogeneous workflow. The GNN learns task dependencies and resource attributes (e.g., cores, memory, features, duration), generating embeddings that encode these relationships. The RL agent, using either these embeddings or the full environment state, assigns tasks to nodes while optimizing for efficiency and feasibility. 

Scheduling constraints—such as resource availability, feature compatibility, and dependency order—are enforced through the reward function. Violations (e.g., insufficient resources, feature mismatches) incur penalties, while efficient task assignments yield positive rewards. This approach allows dynamic updates by modifying the reward function, unlike MILP models that require complete reformulation (\Cref{sec:DetailsOnHandlingMILPConstraintsInGNNRL}).

\textit{GNN-RL models scheduling as a graph:} \textbf{Nodes} represent tasks and compute resources. \textbf{Edges} encode task dependencies and data transfer constraints. The \textbf{GNN} extracts node embeddings for the RL policy. A \textbf{Proximal Policy Optimization (PPO)} \cite{schulmanProximalPolicyOptimization2017} agent assigns tasks, minimizing makespan while satisfying constraints.

Unlike MILP, which struggles with scalability, or heuristics, which oversimplify constraints, GNN-RL learns adaptive policies that generalize across workflows while efficiently handling large-scale HPC scheduling.

\subsubsection{Proposed Method: GrapheonRL – Single Environment with Time-Based Concurrency}

Our proposed method (\Cref{alg:GNNRLPPO}), GrapheonRL, processes large HPC workflows by scheduling tasks in a unified environment that enforces time-based concurrency, dependency checks, and resource constraints. Optionally, the method can leverage imitation data from MILP solutions on smaller sub-DAGs for pre-training. 

\subsubsection{GNN-RL: Algorithm \& Pseudocode}
\label{sec:DetailsOnHandlingMILPConstraintsInGNNRL} 
\begin{algorithm}%[H]
\caption{GNN-RL Scheduler with PPO for HPC Workflows}
\begin{algorithmic}[1]
%\small
\STATE \textbf{Initialize:} Environment $Env$ with tasks $T=\{t_i\}$ and nodes $N=\{n_j\}$, 
policy network $\pi_\theta$ (actor), value function $V_\phi$ (critic), replay buffer $\mathcal{D}$.
\STATE \textbf{Hyperparameters:} learning rate $\alpha$, discount factor $\gamma$, GAE parameter $\lambda$, 
PPO clipping $\epsilon$, PPO epochs $K$, mini-batches $M$, episodes $E$.
\vspace{0.1cm}

\FOR{$episode = 1$ to $E$}
  \STATE \textbf{Initialize} environment state $s_0 \leftarrow Env.reset()$
  \STATE \textbf{Initialize} done $\leftarrow$ \texttt{False}, $t \leftarrow 0$
  \WHILE{\texttt{not done}}
    \STATE \textbf{Compute GNN embeddings:} 
    \STATE \quad $h \leftarrow \mathrm{GNN}\bigl(\mathrm{Graph}(T, \mathrm{dependencies})\bigr)$
    \STATE \textbf{Form augmented state:} $\tilde{s}_t \leftarrow \mathrm{concat}(s_t, h)$
    
    \STATE \textbf{Compute valid action mask:}
    \STATE \quad $\mathcal{M}_t = \{ (t_i, n_j) \mid \text{valid}(t_i, n_j) \}$
    
    \STATE \textbf{Sample action with masking:}
    \STATE \quad $a_t \sim \pi_\theta(a_t \mid \tilde{s}_t, \mathcal{M}_t)$  \COMMENT{(task, node) assignment}
    
    \STATE \textbf{Compute value function:} 
    \STATE \quad $v_t \leftarrow V_\phi(\tilde{s}_t)$
    
    \STATE \textbf{Step environment:} 
    \STATE \quad $s_{t+1}, r_t, \mathrm{done}, \_ \leftarrow Env.step(a_t)$
    
    \STATE \textbf{Store transition:} 
    \STATE \quad $\mathcal{D} \leftarrow (\tilde{s}_t, a_t, r_t, s_{t+1}, v_t, \mathrm{done})$
    \STATE $t \leftarrow t + 1$
  \ENDWHILE

  \STATE \textbf{Compute advantages using GAE:}
  \STATE \quad $\delta_t = r_t + \gamma V_\phi(s_{t+1}) - V_\phi(s_t)$
  \STATE \quad $\hat{A}_t = \sum_{l=0}^{T-t} (\gamma\lambda)^l \delta_{t+l}$
  
  \STATE \textbf{PPO Update:}
  \STATE \quad \textbf{For} $k = 1$ to $K$ \textbf{do:}
  \STATE \quad \quad \textbf{Shuffle and split} $\mathcal{D}$ into $M$ mini-batches
  \STATE \quad \quad \textbf{For each} mini-batch $\mathcal{B}$:
  \STATE \quad \quad \quad \textbf{Compute policy ratio:}
  \[
     r(\theta) = \frac{\pi_\theta(a_t \mid \tilde{s}_t)}{\pi_{\theta_{\mathrm{old}}}(a_t \mid \tilde{s}_t)}
  \]
  \STATE \quad \quad \quad \textbf{Compute clipped surrogate objective:}
  \[
     \mathcal{L}^{\mathrm{CLIP}}(\theta) =
     \mathbb{E}_t \Bigl[
       \min \Bigl(
         r(\theta)\,\hat{A}_t,\;
         \mathrm{clip}\bigl(r(\theta),\,1-\epsilon,\,1+\epsilon\bigr)\,\hat{A}_t
       \Bigr)
     \Bigr]
  \]
  \STATE \quad \quad \quad \textbf{Update actor network:} $\theta \leftarrow \theta + \alpha \,\nabla_\theta \,\mathcal{L}^{\mathrm{CLIP}}(\theta)$
  \STATE \quad \quad \quad \textbf{Update critic network:}
  \[
      \mathcal{L}^{\mathrm{value}}(\phi) =
      \mathbb{E}_t \Bigl[\bigl(G_t - V_\phi(\tilde{s}_t)\bigr)^2\Bigr]
  \]
  \STATE \quad \quad \textbf{end for}
  \STATE \quad \textbf{end for}

  \STATE \textbf{Clear replay buffer $\mathcal{D}$}
  \STATE \textbf{Output} final episode metrics (e.g., total makespan, total reward)
\ENDFOR

\end{algorithmic}
\label{alg:GNNRLPPO}
\end{algorithm}

\paragraph{GNN Embedding}
\[
   h_i = \mathrm{GNN}_{\theta}\bigl(t_i,\;\mathrm{neighbors}(t_i)\bigr),
   \quad\forall\,t_i\in T.
\]

These embeddings can be concatenated into the environment state or used by the policy as extra input features.

\paragraph{PPO Policy Overview} 
We employ PPO to update the RL policy. The process involves computing advantages (e.g., via Generalized Advantage Estimation) and discounted returns \(G_t\) using the value function \(V_\phi(\tilde{s}_t)\), defining and clipping the probability ratio
\[
r(\theta) = \frac{\pi_\theta(a_t \mid \tilde{s}_t)}{\pi_{\theta_{\mathrm{old}}}(a_t \mid \tilde{s}_t)},
\]
and maximizing the clipped surrogate objective
\[
\mathcal{L}^{\mathrm{CLIP}}(\theta) = \mathbb{E}_t\left[\min\left(r(\theta)\,\hat{A}_t,\ \mathrm{clip}\left(r(\theta),\,1-\epsilon,\,1+\epsilon\right)\,\hat{A}_t\right)\right].
\]
Simultaneously, the critic is updated by minimizing the loss
\[
\mathcal{L}^{\mathrm{value}}(\phi) = \mathbb{E}_t\left[\left(G_t - V_\phi(\tilde{s}_t)\right)^2\right].
\]
These steps are iterated over multiple epochs using mini-batches sampled from the collected trajectories.

\subsubsection{Handling Large Workloads}
To enhance scalability, two strategies are considered. The first strategy involves "chunking" the large workflow into smaller DAGs that can be processed concurrently, with results aggregated afterward; however, we found this introduces complexity in synchronizing the chunks along the critical path. Alternatively, in the second strategy, the entire workflow graph is loaded into memory for unified processing. Although this approach requires more memory, it simplifies task ordering and maintains time consistency. In our implementation (\Cref{alg:GNNRLAlgorithm}), we adopt the unified graph approach and store the learned model for subsequent testing.

\begin{algorithm}%[H]
\caption{Unified HPC-CC Environment Scheduling (Single-Environment)}
\begin{algorithmic}[1]
%\small
\STATE \textbf{Definition:}
A set of \textbf{nodes} $\{n_1,\dots,n_k\}$ each with $\{\mathrm{cores}$, $\mathrm{memory}$, $\mathrm{features}\}$.
A set of \textbf{tasks} $\{t_0,\dots,t_{M-1}\}$ each with $\{\mathrm{cores}$, $\mathrm{memory\_required}$, $\mathrm{features}$, $\mathrm{duration}$, $\mathrm{dependencies}\}$.
\textbf{max\_steps}, the episode limit.\\

\STATE \textbf{Initialize} $\mathrm{node\_schedules}[n_j] \leftarrow \emptyset$ for each node $n_j$
\STATE \textbf{Initialize} $\mathrm{global\_assignments} \leftarrow \emptyset$
\STATE \textbf{Initialize} $\mathrm{makespan} \leftarrow 0.0$
\STATE \textbf{Initialize} $\mathrm{state}\in\{0,1\}^{M}$ (all zeros, meaning unassigned tasks)
\STATE \textbf{Initialize} $\mathrm{current\_step}\leftarrow 0$

\WHILE{\textbf{not done}}
    \STATE \emph{RL policy} picks $(\mathrm{task\_idx}, \mathrm{node\_idx})$
    \STATE \(\mathrm{reward}\leftarrow 0.0\)
    \STATE \(\mathrm{current\_step} \leftarrow \mathrm{current\_step}+1\)
    \IF{$\mathrm{current\_step} \ge \mathrm{max\_steps}$}
        \STATE \textbf{Done} $\leftarrow \text{True}$
        \STATE $\mathrm{unassigned\_count}\leftarrow$ count of tasks not in $\mathrm{global\_assignments}$
        \STATE $\mathrm{reward}\leftarrow \mathrm{reward}- (50.0 + 10.0\times \mathrm{unassigned\_count})$
        \STATE \textbf{break}
    \ENDIF

    \STATE \(\mathrm{task\_id} \leftarrow \text{taskKeys}[ \mathrm{task\_idx} ]\)
    \STATE \(\mathrm{node\_id} \leftarrow \text{nodeKeys}[ \mathrm{node\_idx} ]\)

    \STATE \textbf{Check constraints:}
    \begin{enumerate}
       \item \textbf{Already assigned?} If $t_{id}\in \mathrm{global\_assignments}$, $\mathrm{reward}\leftarrow \mathrm{reward}-5$; skip.
       \item \textbf{Dependencies}: If any predecessor $p\notin \mathrm{global\_assignments}$, $\mathrm{reward}\leftarrow \mathrm{reward}-20$; skip.
       \item \textbf{Features}: If $\mathrm{features}(t_{id})\not\subseteq \mathrm{features}(n_{id})$, penalty $-5$; skip.
       \item \textbf{Concurrency}: Find earliest start time $st\ge \max(\text{dep.end\_time})$. Check partial overlap on $n_{id}$’s \emph{node\_schedules} for resource feasibility. 
         If not feasible, $\mathrm{reward}\leftarrow \mathrm{reward}-5$; skip.
    \end{enumerate}

    \STATE \textbf{If valid assignment}:
       \begin{enumerate}
          \item \(\mathrm{start\_time} = st\)
          \item \(\mathrm{end\_time} = st + \mathrm{duration}(t_{id})\)
          \item Append $(\mathrm{start\_time}, \mathrm{end\_time}, \mathrm{cores}, \mathrm{mem})$ to $\mathrm{node\_schedules}[n_{id}]$
          \item $\mathrm{global\_assignments}[t_{id}] \leftarrow (\mathrm{node}=n_{id}, \mathrm{start\_time}, \mathrm{end\_time})$
          \item $\mathrm{makespan} \leftarrow \max(\mathrm{makespan}, \mathrm{end\_time})$
          \item \(\mathrm{state}[ \mathrm{task\_idx} ] \leftarrow 1\)
          \item \textbf{Reward shaping:} $\mathrm{reward} \leftarrow \mathrm{reward} + (15 - \mathrm{duration}+ \mathrm{resource\_factor})$
       \end{enumerate}

    \STATE \textbf{Check if all tasks assigned:}
       \begin{enumerate}
         \item If $|\mathrm{global\_assignments}|= M$, done $\leftarrow \text{True}$
         \item Final bonus: $\mathrm{reward} \leftarrow \mathrm{reward} + (30 - \mathrm{makespan})$
       \end{enumerate}
\ENDWHILE

\STATE \textbf{return} $(\mathrm{global\_assignments}, \mathrm{makespan})$

\end{algorithmic}
\label{alg:GNNRLAlgorithm}
\end{algorithm}

\section{Experimental Setup}
\label{sec:ExperimentalSetup}

\subsection{Test Environment and Datasets}
\label{sssec:TestSystem}
The first test system consist of an edge node (8 cores, 8 GB RAM) connected to the HPC-CC, chosen to assess the solver's capability under limited resources, later we use higher resources for scale test. The solvers are programmed in Python and PuLP \cite{OptimizationPuLPPuLP} library is used for MILP solver.

We use both real and synthetic workflows to experiment various test cases. Real workflows are derived from the Standard Task Graph Set (STGS) \cite{tobita2002standard, kasaharaPracticalMultiprocessorScheduling1984a}, a benchmark for multiprocessor scheduling. Synthetic workflows are generated following programmatically defined workflow pattern for varying node and task counts (see Table~\ref{tab:ScaleTest}). 

\subsection{Test Cases}
\label{sssec:TestCaseDifferentWorkflows}
\begin{figure}
    \centering
    \includegraphics[width=\linewidth]{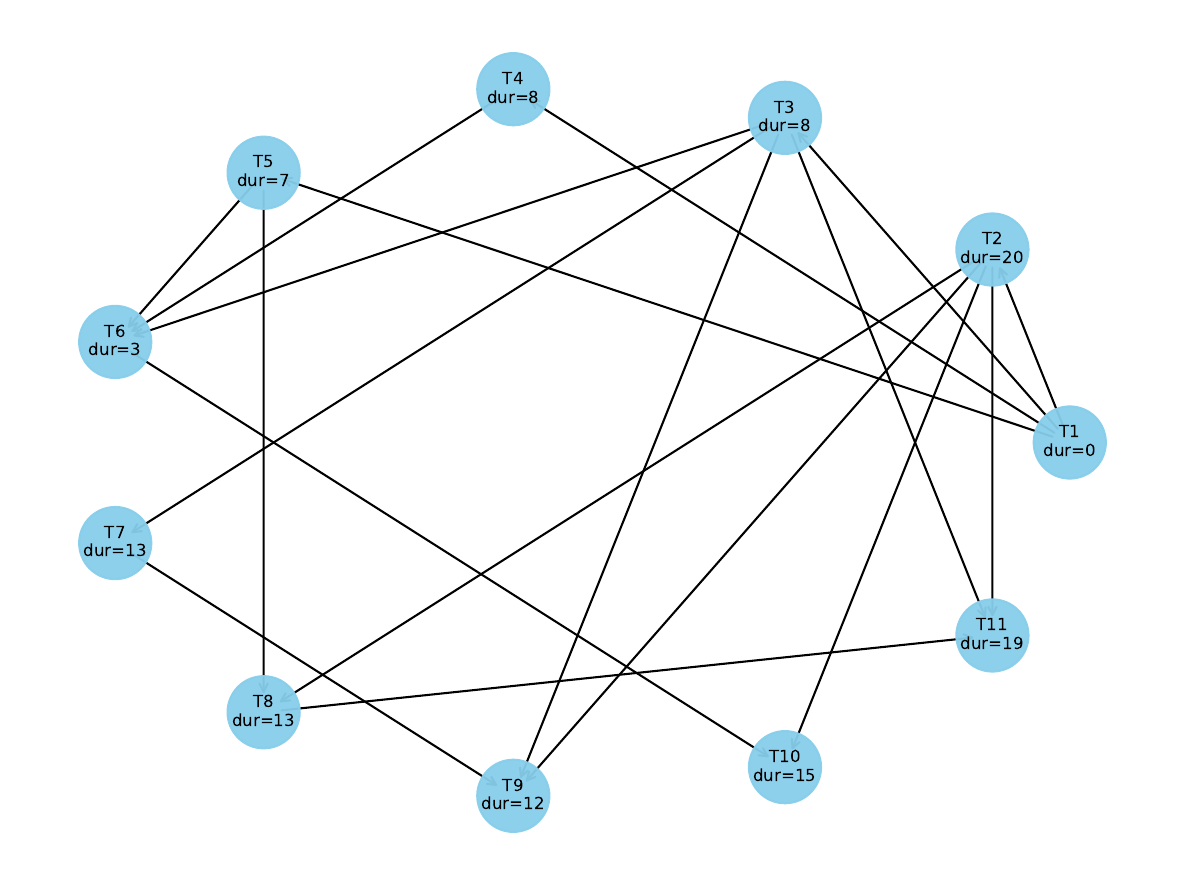} 
    \caption{Workflow 1: Workflow without communication cost}    \label{fig:NoCommunicationCost}
    \vspace{-1em}
\end{figure}

\begin{figure}
    \centering
    \includegraphics[width=\linewidth]{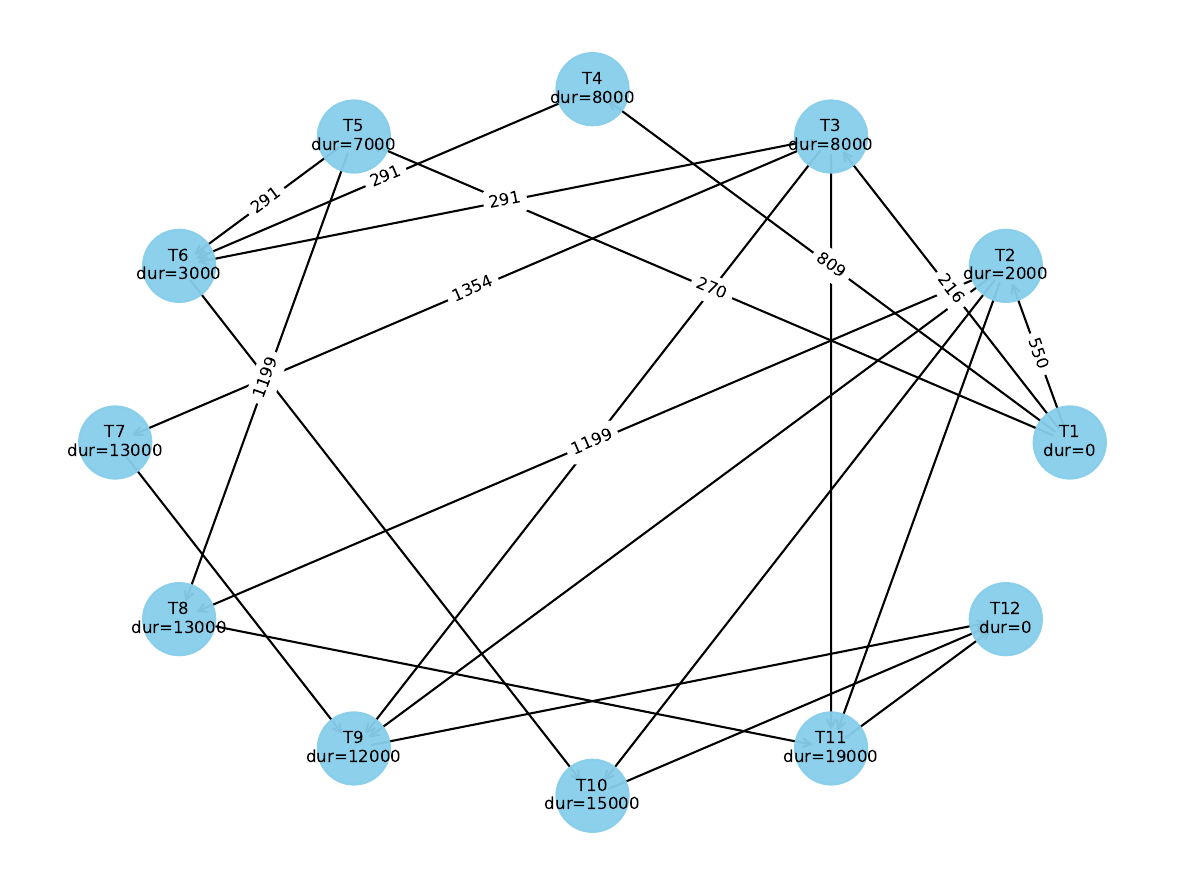} 
    \caption{Workflow 2: Workflow with communication cost}    \label{fig:WithCommunicationCost}
    \vspace{-1em}
\end{figure}

\begin{figure}
    \centering
    \includegraphics[width=\linewidth]{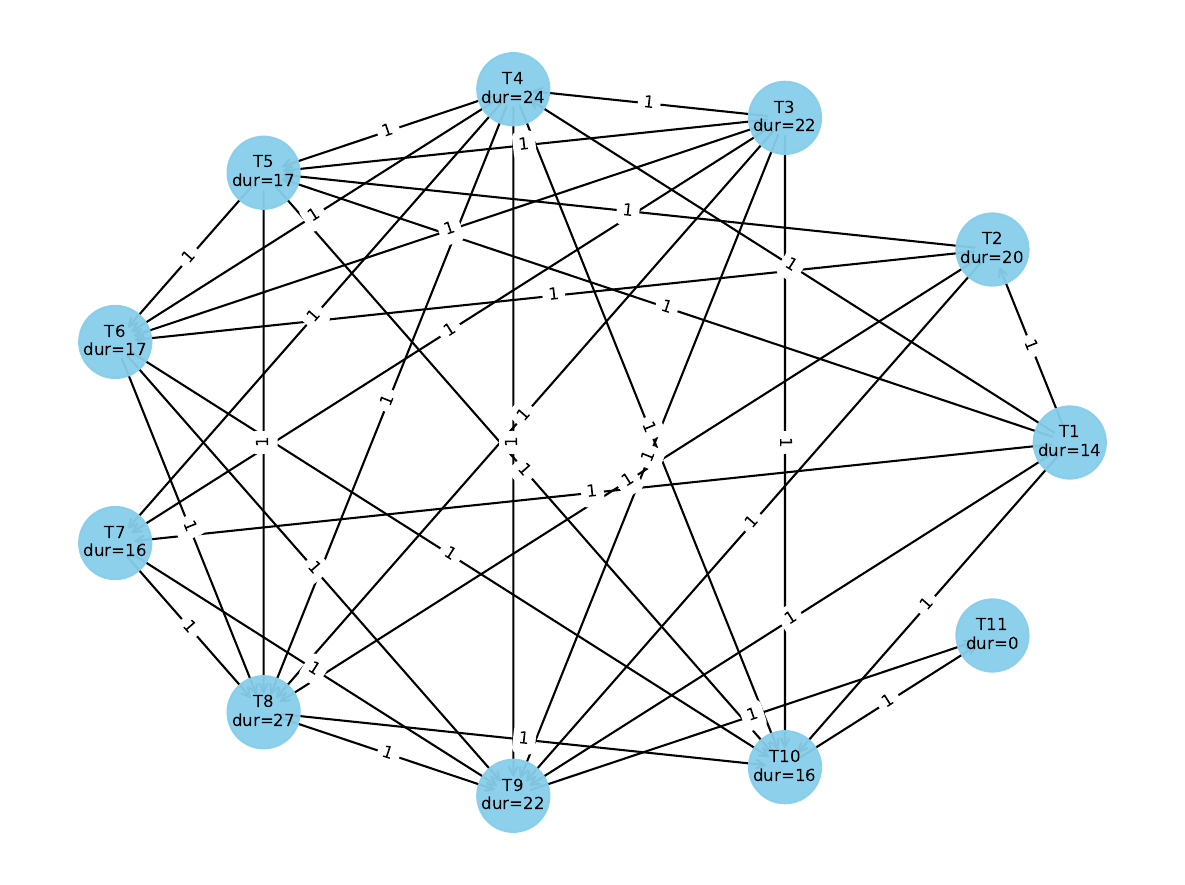} 
    \caption{Workflow 3: Tasks with higher level of connecting edges and with a default communication cost.}
    \label{fig:PrototypeNoCommCost}
    \vspace{-1em}
\end{figure}

\begin{figure}
    \centering
    \includegraphics[width=\linewidth]{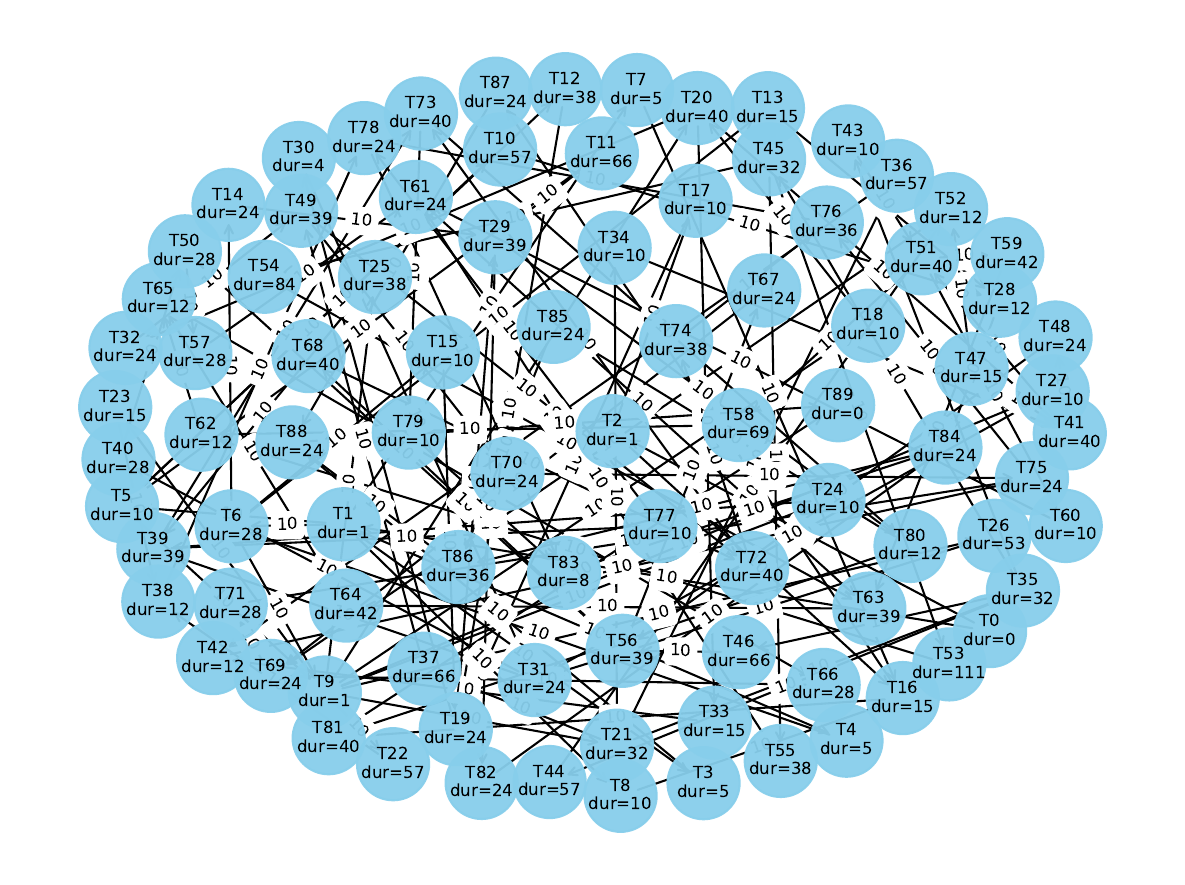} 
    \caption{Workflow 4: Robot control STGS (complex) graph.}
    \label{fig:RobotControlSTG}
    \vspace{-1em}
\end{figure}

We evaluate four real workflows derived from STGS:
\textit{Workflow 1:} No data transfer time (DTT) (\Cref{fig:NoCommunicationCost}).
 \textit{Workflow 2:} With high DTT per task (\Cref{fig:WithCommunicationCost}).
 \textit{Workflow 3:} Default DTT combined with increased task dependencies (\Cref{fig:PrototypeNoCommCost}).
 \textit{Workflow 4:} A complex Robot Control STGS graph (\Cref{fig:RobotControlSTG}).

Each workflow assigns time unit per task, with designated source and sink tasks. Standardized input characteristics are shown in \Cref{sssec:SystemAndWorkloadInputFormat}, whereas each workflow 1, 2, 3, and 4 has the task counts of 10, 12, 11, and 89, respectively. The workflows are further processed against techniques like the MILP, GNN-RL framework, and Heuristics (OLB and HEFT).

\subsubsection{System and Workload Input Characteristics}
\label{sssec:SystemAndWorkloadInputFormat}
\begin{figure}[ht!]
    \vspace{-1em}
    \small  
    \begin{verbatim}      
{
    "nodes": {
        "node1": {
            "cores": 32,
            "memory": 15742900633600.0,
            "features": ["F1"],
            "processing_speed": 2.5,
            "data_transfer_rate": 100
        },
        ....
    }
}
    \end{verbatim}
    \normalsize
    \caption{Cluster nodes input format in JSON.}
    \label{fig:NodeInputCharacteristics}
    \vspace{-1em}
\end{figure}

\begin{figure}[ht!]
    \vspace{-1em}
    \small  
    \begin{verbatim}      
{
    "workflows": {
        "Workflow1": {
            "tasks": {
                "T1": {
                    "cores": 10,
                    "memory_required": 20,
                    "features": ["F1", "F2"],
                    "data": 10,
                    "duration": [100, 50],
                    "dependencies": []
                },
                ....
            }
        }
    }
}
    \end{verbatim}
    \normalsize
    \caption{Workflow input format in JSON.}
    \label{fig:WorkloadInputCharacteristics}
    \vspace{-1em}
\end{figure}

\subsection{Evaluation Metrics}
\subsubsection{Performance}
Performance is assessed based on - \textit{Execution Time and Makespan:} Measured in seconds, \textit{Memory Utilization:} Recorded in MB, and \textit{Task-to-Node Mapping Efficiency:} Evaluated by comparing the number of tasks to available nodes, respectively. %For the performance evaluation we used three nodes with heterogeneous resources (16, 32 and 48 cores). These nodes resembles with nodes configuration of "Anonymous Compute Cluster" having 1771 nodes that belongs to Anonymous. 
\begin{table*}
    \centering
    \caption{Performance Details}
    \begin{tabular}{cp{12em}lrrrrr}
    %{c p{5em}p{7em}p{3em}p{3em}p{3em}p{3em}p{3em}p{8em}}
        \hline
        \textbf{Workflow} & \textbf{Workflow Type} & \textbf{Method} & \textbf{Num Nodes} & \textbf{Num Tasks} & \textbf{Makespan} & \textbf{Solver Time (s)} & \textbf{Mem Diff (MB)} \\
        \hline
          &  & MILP & 3 & 11 & 52 & 0.02 & 0.20 \\
         &  & GNN-RL-Train-Test & 3 & 11 & 52 & 139.41 & 103.10 \\
        1 & No Communication Cost & GNN-RL-Test & 3 & 11 & 52 & 0.01 & 0.10 \\
         &  & OLB & 3 & 11 & 60 & 0.01 & 0.01 \\
         &  & HEFT & 3 & 11 & 43 & 0.01 & 0.01 \\
         \hline
          &  & MILP & 3 & 12 & 39 & 0.04 & 0.20 \\
         &  & GNN-RL-Train-Test & 3 & 12 & 39 & 143.12 & 102.90 \\
        2 & With Communication Cost* & GNN-RL-Test & 3 & 12 & 39 & 0.04 & 0.10 \\
         &  & OLB & 3 & 12 & 42 & 0.01 & 0.01 \\
         &  & HEFT & 3 & 12 & 35 & 0.01 & 0.01 \\
         \hline
          &   & MILP & 3 & 11 & 129 & 0.08 & 0.40 \\
         &  & GNN-RL-Train-Test & 3 & 11 & 129 & 140.81 & 102.90 \\
        3 & No Communication Cost & GNN-RL-Test & 3 & 11 & 129 & 0.06 & 0.10 \\
         & High Connections & OLB & 3 & 11 & 129 & 0.05 & 0.04 \\
         &  & HEFT & 3 & 11 & 71 & 0.04 & 0.03 \\
         \hline
          &  & MILP & 3 & 90 & 569 & 0.43 & 1.5 \\
         &  & GNN-RL-Train-Test & 3 & 90 & 569 & 176.02 & 101.40 \\
        4 & Robot Control STG & GNN-RL-Test & 3 & 90 & 569 & 1.81 & 0.10 \\
         &  & OLB & 3 & 90 & 1160 & 0.08 & 0.06 \\
         &  & HEFT & 3 & 90 & 829 & 0.09 & 0.05 \\
        \hline
        \multicolumn{7}{l}{*\tiny To adjust the result with other graph outcome, the communication time cost is converted from seconds to milliseconds.}
    \end{tabular}
    \label{tab:PerformanceDetails}
\end{table*}

The results confirm GNN-RL as a scalable alternative to MILP for HPC scheduling. While MILP achieves optimal makespan, its solver time grows exponentially, reaching $0.43s$ for $90$ tasks. In contrast, GNN-RL (Test phase) matches MILP’s makespan ($569$) in $1.81s$—$76\%$ faster in inference. GNN-RL (Train-Test phase) incurs a $176.02s$ training cost but generalizes across workflows for rapid scheduling. HEFT outperforms OLB by $16$–$30\%$, especially in high-dependency cases where OLB's makespan is $103.8\%$ worse than MILP (e.g., $1160$ vs. $569$). For small tasks, MILP remains feasible ($<0.1s$ solver time), but for larger workflows, GNN-RL achieves near-optimal solutions with a negligible $0.10$ MB memory footprint vs. MILP’s $1.5$ MB, making it ideal for large-scale HPC environments.

\subsubsection{Makespan: Overall completion time} 
\begin{figure}%[H]
    \centering
    \includegraphics[width=\linewidth]{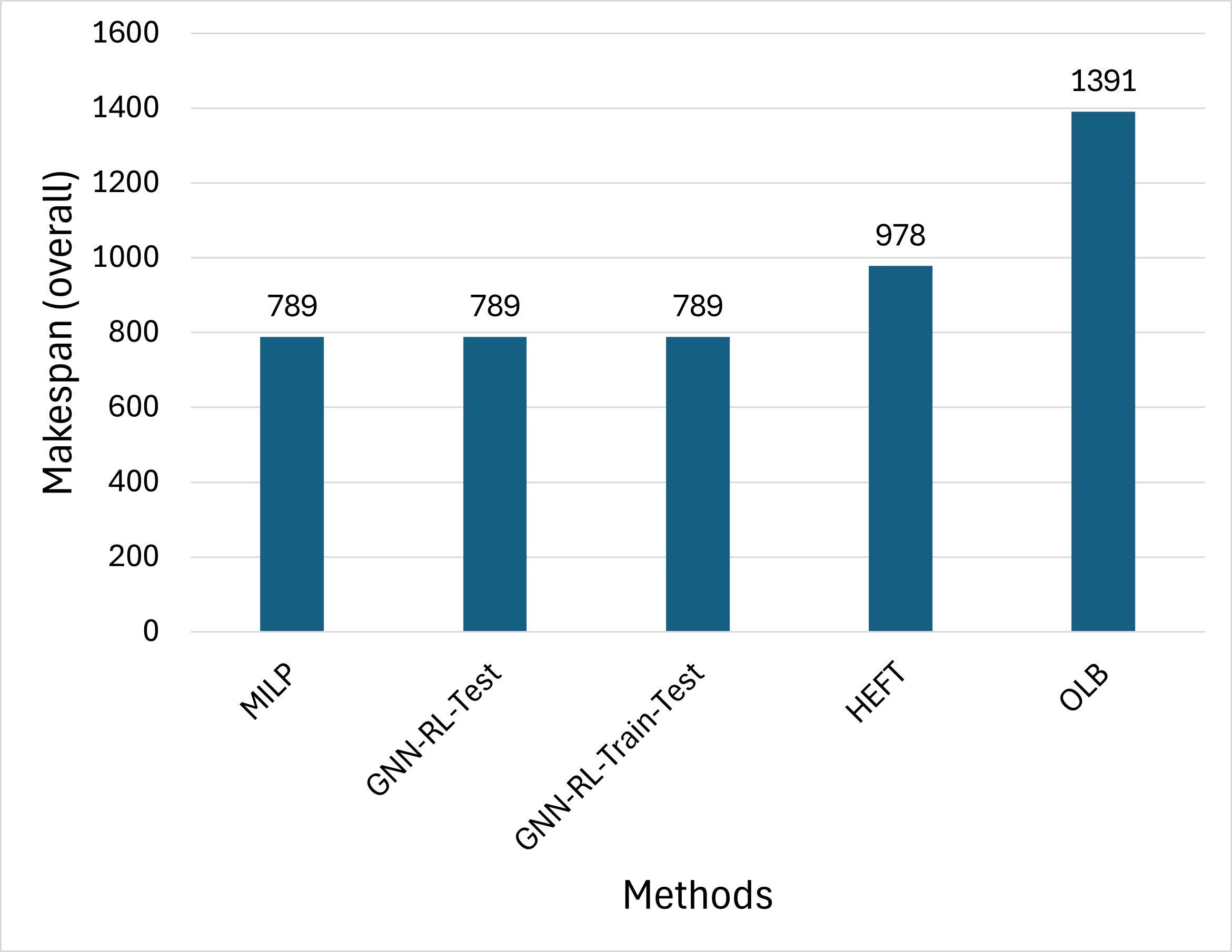}
    \caption{Overall makespan of methods on different workflows based on \Cref{tab:PerformanceDetails}.}
    \label{fig:OverallMakespan}
\end{figure}
As our main object is to minimize makespan by optimally utilizing the available resources. In this regards, the result shown in \Cref{fig:OverallMakespan} supports that GrapheonRL is capable to provide optimum makespan based on the given resources similar to MILP in overall, as the result matches with the minimum makespan given by MILP solution.

\subsubsection{Scalability} 
In regards to the impact of scaling workflow size. For this purpose we utilized a synthetic nodes and workflows generator with a pattern using an arbitrary values because our main objective in this test is to verify the workload size that the solver is capable to handle. 
\begin{table}%[H]
    \centering
    \caption{Scale Test}
    \begin{tabular}{clrcr}
        \hline
        \textbf{Num} & \textbf{Method} & \textbf{Makespan} & \textbf{Resource} & \textbf{Runtime} \\
        \textbf{Nodes x Tasks} & \textbf{} & \textbf{} & \textbf{Used} & \textbf{(seconds)} \\
        
        \hline
        10 x 10 & GNN-RL-Test & 41 & Optimally & 0.65\\
        10 x 10 & OLB & 41 & Optimally & 0.01\\
        10 x 10 & HEFT & 10 & Greedily & 0.01\\
        \hline
        100 x 100 & GNN-RL-Test & 60 & Optimally & 1.58\\
        100 x 100 & OLB & 60 & Optimally & 0.09\\
        100 x 100 & HEFT & 10 & Greedily & 0.11\\
        \hline
        1000 x 1000 & GNN-RL-Test & 95 & Optimally & 27.78\\
        1000 x 1000 & OLB & 95 & Optimally & 0.72\\
        1000 x 1000 & HEFT & 10 & Greedily & 0.93\\
        \hline
        10000 x 10000 & GNN-RL-Test & 189 & Optimally & 290.40\\
        10000 x 10000 & OLB & 189 & Optimally & 75.51\\
        10000 x 10000 & HEFT & 10 & Greedily & 107.86\\
        \hline
    \end{tabular}
    \label{tab:ScaleTest}
\end{table}

%- \textit{Adaptability} (ease of updating constraints/objectives).\\

As shown in \Cref{tab:ScaleTest}, Scale tests show GNN-RL maintains optimality across problem sizes but incurs higher computation costs. At $10 \times 10$ tasks, all methods achieve a makespan of $41$, with HEFT being the fastest ($0.01s$). However, HEFT fails at scale, maintaining an inefficient makespan of $10$, while GNN-RL and OLB remain optimal ($60$ at $100 \times 100$, $95$ at $1000 \times 1000$, $189$ at $10,000 \times 10,000$). GNN-RL’s runtime grows from $1.58s$ ($100$ tasks) to $290.40s$ ($10,000$ tasks), making it $3.85\times$ slower than OLB at large scales. OLB remains fastest but lacks adaptability, while HEFT is $94.7$\% worse than optimal for large workflows, proving unsuitable for large-scale scheduling. These results establish GNN-RL as an effective AI-based scheduler, balancing optimality and scalability, though at a computational cost.

\section{Analysis and Discussion}
\label{sec:ResultsAndDiscussion}

\subsection{Evaluation of Experiments}
MILP achieves optimal scheduling for small workloads but scales poorly. Heuristics like HEFT and OLB provide fast approximations but struggle with complex constraints. Our \textit{GNN-RL scheduler balances scalability, adaptability, and efficiency}, dynamically adjusting via reward function modifications while efficiently handling dependencies.

\subsubsection{Algorithmic Complexity}
The GNN-RL model has an action space complexity of \(O(M \times N)\), where \(M\) tasks are mapped to \(N\) nodes. Time-based concurrency verification is \(O(M \times N)\), and RL training over \(S\) steps results in \(O(S \cdot N)\) total complexity. Despite higher training overhead, inference is efficient and scales well.

\subsubsection{Performance Comparison}
\textbf{MILP vs. GNN-RL:} MILP is optimal for $\leq 12$ tasks but struggles with larger workflows, whereas GNN-RL maintains near-optimal makespan even at $90+$ tasks. Training is expensive (up to $176s$), but inference is efficient ($~1.81s$).

\textbf{Heuristics vs. GNN-RL:} HEFT and OLB execute quickly ($\leq0.09s$) but yield suboptimal schedules ($94.7$\% worse in some cases, e.g., $1160$ vs. $569$ makespan for Robot Control STG). HEFT fails at 10,000 tasks, maintaining a static makespan of 10, whereas GNN-RL dynamically optimizes scheduling (makespan = 189).

\subsubsection{Scalability and Real-Time Suitability}
GNN-RL effectively schedules 10,000-task workloads, unlike heuristics, but incurs a runtime penalty (290.40s vs. 75.51s for OLB). Future optimizations are needed for real-time performance.

\subsubsection{Applicability}
MILP is best for small-scale deterministic scheduling, heuristics for rapid approximations, and GNN-RL for large-scale dynamic HPC workloads, particularly where adaptability to evolving constraints is required.

\section{Conclusion and Future Work}
\label{sec:ConclusionAndFutureWork}

GNN-RL effectively models resource concurrency, dependencies, and scheduling constraints, offering a \textit{scalable alternative to MILP and heuristics}. While training is costly, inference is fast ($76$\% faster then MILP and only $3.85\times$ slower then OLB), making it suitable for large, adaptive workloads.
Future directions include advanced GNN architectures, and online RL for real-time scheduling, along with distributed/hierarchical approaches for extreme-scale workflows (100K+ tasks).

Note: the project codes will be available through Github once the proceeding are completed.

\section*{Acknowledgments}
\label{sec:Acknowledgments}

This work is supported by the European Horizon Project DECICE (Grant No. 101092582) and the National High Performance Computing (NHR: (visible at: \url{www.nhr-verein.de/unsere-partner})) initiative. We thank GWDG for their significant contributions to this research.

The implementation source code is available at \url{https://github.com/AasishKumarSharma/GrapheonRL}.

\bibliographystyle{IEEEtran} 
\bibliography{bibliotheke}

\end{document}